\documentclass[fleqn,twoside]{article}
\usepackage{espcrc2}
\input epsf.tex

\usepackage{graphicx}
\usepackage[figuresright]{rotating}

\def\half{\frac{1}{2}}
\def\quarter{\frac{1}{4}}
\def\third{\frac{1}{3}}

\newcommand{\AmS}{{\protect\the\textfont2
  A\kern-.1667em\lower.5ex\hbox{M}\kern-.125emS}}

\title{ Neutrino properties from high energy astrophysical neutrinos}

\author{Sandip Pakvasa\address{Department of Physics and Astronomy, University of Hawaii, 
        Honolulu, HI  96822,  USA} \\
}
\begin{document}

\begin{abstract}
It is shown how high energy neutrino beams from 
very distant sources can be utilized to learn about some properties
of neutrinos such as lifetimes, mass hierarchy, etc.  Furthermore, 
even mixing elements such as $U_{e3}$ and the CPV phase in the 
neutrino mixing matrix can be measured
in principle.  Pseudo-Dirac mass differences as small
as $10^{-18} eV^2$ can be probed as well.
\vspace{1pc}
\end{abstract}

\maketitle

\normalsize\baselineskip=15pt

\section{Introduction}
We make two basic assumptions which are reasonable.  The first one is that distant neutrino
sources (e.g. AGN's and GRB's) exist; and furthermore with detectable fluxes
at high energies (upto and beyond PeV).  The second one is that in the not
too far future, very large volume, well instrumented detectors of sizes of
order of KM3 and beyond will exist and be operating; and furthermore will
have (a) reasonably good energy resolution and (b) good angular resolution
($\sim 1^0 $ for muons).

\section{Neutrinos from Astrophysical Sources}

If these two assumptions are valid, then there are a number of uses these
detectors can be put to\cite{pakvasa}.  In this talk I want to focus on those that enable us
to determine some properties of neutrinos: namely, probe neutrino lifetimes 
to $10^4 s/eV$ (an improvement of $10^8$ over current bounds), 
pseudo-Dirac mass splittings to a level of $10^{-18} eV^2$ (an improvement of a factor of $10^6$ over
current bounds) and potentially even measure quantities such as $U_{e3}$ and the phase $\delta$ in the MNSP matrix\cite{MNS}.

\section{Astrophysical neutrino flavor content}

In the absence of neutrino oscillations we expect a very small $\nu_\tau$
component
in neutrinos from astrophysical sources. From the most discussed and  the most likely astrophysical high energy
neutrino sources\cite{learned} we expect nearly equal numbers of particles
and anti-particles, half as many $\nu_e's$ as $\nu_\mu's$ and virtually no $\nu_\tau's$.  This comes about simply 
because the neutrinos are thought to originate in decays of pions (and
kaons) and subsequent decays of muons.  Most astrophysical targets are fairly tenous even compared to the Earth's atmosphere, and would allow for full muon decay  
in flight.  There are some predictions for flavor independent fluxes from cosmic defects and exotic objects such as evaporating black holes.  Observation of
tau neutrinos from these would have great importance.  A conservative estimate\cite{learned1} shows that the prompt $\nu_\tau$ flux is very small and the emitted flux is close to the ratio $1:2:0$.  The flux ratio of $\nu_e: \nu_\mu: \nu_\tau = 1:2:0$
is certainly valid for those AGN models in which the neutrinos are 
produced in beam dumps of photons or protons on matter, in which
mostly pion and kaon decay(followed by the decay of muons) supply the bulk of 
the neutrino flux. 

Depending on the amount of prompt $\nu-$flux due to the production and decay
of heavy flavors, there could be a small non-zero $\nu_\tau$ component 
present. There are also possible scenarios in which the muons lose energy in 
matter or
in strong magnetic fields\cite{rachen}, in which case the initial flux mixture becomes
$\nu_e:\nu_\mu:\nu_\tau = 0:1:0$.  

\section{Effect of Oscillations}

The current knowledge of neutrino masses and mixings can be summarized 
as follows\cite{pakvasa1}. The mixing matrix is given to a good approximation by 
\begin{eqnarray}
U = \left (
\begin{array}{ccc}
c & s & \epsilon \\
s / \sqrt{2}  & c / \sqrt{2} & 1/ \sqrt{2} \\
s / \sqrt{2} & c / \sqrt{2} & 1/ \sqrt{2}
\end{array} \right )
\end{eqnarray}
where $c=\cos \theta, s= \sin \theta$ with $\theta$ the solar mixing angle
given by about $32^0$ and $\epsilon = U_{e3} < 0.17$ limited by the CHOOZ bound.  The mass spectrum has two possibilities; normal or inverted, and 
with the mass differences given by $\delta m^2_{32} \sim 2.10^{-3} eV^2$ and 
$\delta m_{21}^2 \sim 7.10^{-5} eV^2$.  Since $\delta m^2 L/4E$ for 
the distances to GRB's and AGN's (even for energies upto and beyond PeV) 
is very large $(> 10^7)$ the oscillations have always averaged out 
and the conversion and survival probabilities are given by
\begin{eqnarray}
P_{\alpha \beta} &=& \sum_{i} | U_{\alpha i} \mid^2 \mid U_{\beta i} \mid^2 \\
P_{\alpha \alpha} &=& \sum_{i} \mid U_{\alpha i} \mid^4
\end{eqnarray}
Assuming no significant matter effects enroute, 
the mixing matrix in Eq. (1) leads to a propagation matrix P, given by:
\begin{eqnarray}
P = \left(
\begin{array}{ccc}
1- S/2  &  S/4            &  S/4   \\
S/4   &  1/2 - S/8      &  1/2-S/8  \\
S/4    &  1/2 - S/8      &  1/2-S/8  \\
\end{array} \right)
\end{eqnarray}
where S stands for $sin^2(2\theta)$. As is obvious, for any value of the solar mixing angle, P converts a flux ratio of $\nu_e: \nu_\mu: \nu_\tau = 1:2:0$ 
into one of $1:1:1$.  Hence the flavor mix expected at arrival 
is simply an equal mixture of $\nu_e, \nu_\mu$ and 
$\nu_\tau$ as was observed long ago\cite{learned1,athar}.
An initial flavor mix of $\nu_e:\nu_\mu:\nu_\tau = 0:1:0$ is converted by
oscillations into one of about  $1/2:1:1$.
There are several other ways, arising from intrinsic properties of neutrinos, 
by which the flavor mix can change from the canonical $1:1:1$ figure. One in particular, which gives rise to striking signatures, is the decay 
of neutrinos\cite{beacom}.  Before discussing the other possibilities, let me
consider the case of neutrino decay. 

\section{Neutrino Decay\cite{pakvasa2}}

We now know that neutrinos have non-zero masses and non-trivial mixings,
based on the evidence for neutrino mixings and oscillations from the data on
atmospheric, solar and reactor neutrinos.

If this is true, then in general, the heavier neutrinos are expected to decay into the lighter ones via flavor changing processes.  The only questions are (a) whether the lifetimes are short enough to be phenomenologically interesting (or are they too long?) and (b) what are the dominant decay modes.

Throughout the following, to be specific, I will assume that the
neutrino masses are at most of order of eV.  Since we are interested
in decay modes which are likely to have rates (or lead to lifetimes)  which are phenomenologically interesting,  we can rule out several classes of decay nodes.

First, consider radiative decays, such as $\nu_i \rightarrow \nu_j 
+ \gamma$.  Since the experimental bounds on $\mu_{\nu _i}$, the magnetic
moments of neutrinos, come from reactions such as 
$\nu_e e \rightarrow e``\nu''$ which are not sensitive to the final state neutrinos; the bounds
apply to both diagonal as well as transition magnetic moments and so can be
used to limit the corresponding lifetimes. The bounds should really be on
mass eigenstates\cite{john}, but since the mixing angles are large, it does not matter
much.  The current bounds are\cite{these}:
\begin{eqnarray}
\tau_{\nu_ e}  > & 5.10^{18} \ \mbox{sec}  \nonumber \\
\tau_{\nu_\mu}   > & 5.10^{16} \ \mbox{sec} \\
\tau_{\nu_\tau}  > & 2.10^{11} \ \mbox{sec} \nonumber 
\end{eqnarray}
In the above limits the first one gives a bound for the $\tau_{\nu_1}$,
whereas the second one gives the bound for both $\tau_{\nu_2}$ as well
as $\tau_{\nu_3}$ since the mixing is essentially maximal.

There is one caveat in deducing these bounds.  Namely, the form factors are
evaluated at $q^2 \sim O (eV^2)$ in the decay matrix elements whereas in the scattering from which the bounds are derived, they are evaluated 
at $q^2 \sim O (MeV^2)$.  Thus, some extrapolation is necessary.  It can be
argued that, barring some bizarre behaviour, this is justified\cite{frere}.

An invisible decay mode with no new particles is the three body decay 
$\nu_i \rightarrow \nu_j \nu_j \bar{\nu}_j$.  Even at the full 
strength of Z coupling, this yields a lifetime of $2.10^{34}$s, far too long to be of interest.  There is an indirect bound from Z decays which is weaker but still yields 2.10$^{30}$s \cite{bilenky}.

Thus, the only decay modes which can have interestingly fast decays rates are two body modes such as $\nu_i \rightarrow \nu_j + x$ and $\nu_i \rightarrow \bar{\nu}_j + x$ where $x$ is a very light or massless particle, e.g. a Majoron.

The only possibility for fast invisible decays of neutrinos seems to lie 
with Majoron or Majoron-like models\cite{pakvasa2}.  There are two classes of models; the I=1 Gelmini-Roncadelli\cite{gelmini} majoron and the I=0 Chikasige-Mohapatra-Peccei\cite{chicasige} majoron. In general, one can choose the majoron to be a mixture of the two; furthermore the coupling can be to flavor as well as sterile neutrinos.  The effective interaction is of the form:
\begin{equation}
\bar{\nu}^c_\beta (a+b \gamma_5) \nu_\alpha \ J
\end{equation}
giving rise to decay:
\begin{equation}
\nu_\alpha \rightarrow \bar{\nu}_\beta \ ( or  \ \nu_\beta)  +  J
\end{equation}

where $J$ is a massless $J= 0 , L =2$ particle; $\nu_\alpha$ and $\nu_\beta$
are mass eigenstates which may be mixtures of flavor and sterile neutrinos.
Models of this kind which can give rise to fast neutrino decays have been discussed\cite{valle}.  These models are unconstrained by $\mu$ and $\tau$ decays which do not arise due to the $\Delta L = 2$ nature of the coupling.  The I=1 coupling is constrained by the bound on the invisible
$Z$ width; and requires that the Majoron be a mixture of I=1 and
I=0\cite{choi}.
 The couplings of $\nu_\mu$ and $\nu_e$ $(g_\mu$ and $g_e)$ are constrained
by the limits on multi-body $\pi$, K decays
$\pi \rightarrow \mu \nu \nu \nu$ and $K \rightarrow \mu \nu \nu \nu$ and
on $\mu-e$ university violation in $\pi$ and K decays\cite{barger}, but not
sufficiently strongly to rule out fast decays.

There are very interesting cosmological implications of such couplings. The
details depend on the spectrum of neutrinos and the scalars in the model.
For example, if all the neutrinos are heavier than the scalar; the relic
neutrino density vanishes today, and the neutrino mass bounds from CMB
and large scale structure are no longer operative, whereas future
measurements in the laboratory might find a non-zero result for a neutrino
mass \cite{beacom1}. 
If the scalars are heavier than the neutrinos, there are signatures
such as shifts of the $n$th multipole peak (for large $n)$ in the 
CMB \cite{chacko}.
There are other implications as well, such as the number of relativistic
degrees of freedom(or effective number of neutrinos) being different at the
BBN and the CMB eras. The additional degrees of freedom should be detectable
in future CMB measurements.

Direct limits on such decay modes are also very weak.
Current bounds on such decay modes are as follows.  For the mass eigenstate $\nu_1$, the limit is about
\begin{equation}
\tau_1 \geq 10^5 \ sec /eV
\end{equation}
based on observation of $\bar{\nu}_e's$ from SN1987A \cite{hirata}
(assuming CPT invariance). For $\nu_2$, strong  limits can be deduced  from
the non-observation of solar anti-neutrinos in KamLAND\cite{eguchi}  but
only in the case when the coupling is to $\nu_1$. In the most general case,
an analysis of solar neutrino data\cite{bell} leads to a bound given by:
\begin{equation}
\tau_2 \geq 10^{-4} \ sec/eV
\end{equation}
For $\nu_3$, in case of  normal hierarchy, one can derive a bound from the atmospheric neutrino observations of upcoming neutrinos\cite{barger1}:
\begin{equation}
\tau_3 \geq \ 10^{-10} \ sec/eV
\end{equation}

The strongest lifetime limit is thus too weak to eliminate the possibility of
astrophysical neutrino decay by a factor about $10^7 \times (L/100$ Mpc) 
$\times (10$ TeV/E).  Some aspects of the decay of high-energy astrophysical
neutrinos have been considered in the past.  It has been noted that the
disappearance of all states except $\nu_1$ would prepare a beam that could in principle be used to measure elements of the neutrino mixing matrix, namely the ratios $U^2_{e1} : U^2_{\mu 1} : U^2_{\tau 1}$\cite{pakvasa3}.  
The possibility of measuring
neutrino lifetimes over long baselines was mentioned in Ref.\cite{weiler}, and some predictions for decay in four-neutrino models were given in Ref.\cite{keranen}.  We have shown that the particular values and small uncertainties on the neutrino mixing parameters allow for the first time very distinctive signatures of the effects of neutrino decay on the detected flavor ratios.  The expected increase in neutrino lifetime sensitivity (and corresponding anomalous 
neutrino couplings) by several orders of magnitude makes for a very
interesting test of physics beyond the Standard Model; a discovery would
mean physics much more exotic than neutrino mass and mixing alone.  As shown
below,  neutrino decay because of its unique signature cannot be mimicked by either different neutrino flavor ratios at the source or other non-standard neutrino interactions.

A characteristic feature of decay is its strong energy dependence: 
$\exp (-Lm/E \tau)$, where $\tau$ is the rest-frame lifetime.  
For simplicity, we will assume that decays are always complete, i.e., that 
these exponential factors vanish.  The assumption of complete decay means we do not have to consider the distance and intensity distributions of sources.  We assume an isotropic diffuse flux of high-energy astrophysical neutrinos, and can thus neglect the angular deflection of daughter neutrinos from the trajectories of their parents\cite{lindner}.

{\bf Disappearance only.-}  Consider the case of no detectable decay
products, that is, the neutrinos simply disappear.  This limit is interesting for decay to 'invisible'  daughters, such as a sterile neutrino, and also for decay
to active daughters if the source spectrum falls sufficiently steeply with
energy.  In the latter case, the flux of daughters of degraded energy will
make a negligible contribution to the total flux at a given energy.  Since coherence will be lost we have.
\begin{eqnarray}
\phi_{\nu_{\alpha}} = 
\sum_{i \beta} \phi^{source}_{\nu _\beta} (E) 
\mid U_{\beta i} \mid^2 \mid U_{\alpha i} \mid^2 e^{-L/\tau_i (E)} \\
 \stackrel{L \gg \tau_i}{\longrightarrow}  
\sum_{i(stable), \beta}  \phi^{source}_{\nu_{\beta}} (E) \mid U_{\beta i} \mid^2 \mid U_{\alpha i} 
\mid^2 ,
\end{eqnarray}
where the $\phi_{\nu_{\alpha}}$ are the fluxes of $\nu_\alpha, U_{\alpha i}$
are elements of the neutrino mixing matrix and $\tau$ are the neutrino
lifetimes in the laboratory frame.  Eq. (5) corresponds to the case where
decay is complete by the time the neutrinos reach us, so only the stable 
states are included in the sum.

The simplest case (and the most generic expectation) is a normal hierarchy 
in which both $\nu_3$ and $\nu_2$ decay, leaving only the 
lightest stable eigenstate  $\nu_1$.  In this case the 
flavor ratio is 
$U^2_{e1}:  U^2_{\mu 1} : U^2_{\tau 1}$\cite{pakvasa3}. 
Thus if $U_{e3} = 0$
\begin{equation}
\phi_{\nu e} :  \phi_{\nu_{\mu}} :  \phi_{\nu_{\tau}}
\simeq 5 : 1 : 1, 
\end{equation}
where we used the neutrino mixing parameters given above\cite{beacom}.  
Note that this is an extreme deviation of the flavor ratio from
that in the absence of decays.  It is difficult to imagine other mechanisms
that would lead to such a high ratio of $\nu_e$ to $\nu_\mu$.  In the case
of inverted hierarchy, $\nu_3$ is the lightest and hence stable state, and
so\cite{beacom}
\begin{equation}
\phi_{\nu_{e}} :  \phi_{\nu_{\mu}} : \phi_{\nu _{\tau}} = U^2_{e3} : 
U^2_{\mu 3} : U^2_{\tau 3} = 0 : 1 : 1.
\end{equation}
If  $U_{e3} = 0$ and $\theta_{atm} = 45^0$, each mass eigenstate has equal
$\nu_\mu$ and $\nu_\tau$ components.  Therefore, decay cannot break 
the equality between the $\phi_{\nu_{\mu}}$ and $\phi_{\nu_{\tau}}$ 
fluxes and thus the $\phi_{\nu_{e}} : \phi_{\nu_\mu}$ ratio contains all the useful information. 
The effect of a non-zero $U_{e3}$ on the no-decay case of 1 : 1 : 1 
is negligible.

When $U_{e3}$ is not zero, and the hierarchy is normal, it is possible to
obtain information on the values of $U_{e3}$ as well as the CPV phase $\delta$\cite{beacom2}.  The flavor ratio $e/\mu$ varies from 5 to 15 (as $U_{e3}$ goes from 0 to 0.2)  
for $\cos \delta =+1$ but from 5 to 3 for $\cos \delta =-1$.  The ratio $\tau/\mu$ varies from 1 to 5 $(\cos \delta = +1)$ or 1 to 0.2 $(\cos \delta =-1)$ for the same range of $U_{e3}$.

If the decays are not complete and if the daughter does not carry the full
energy of the parent neutrino; the resulting flavor mix is somewhat
different but any case it is still quite distinct from the simple $1:1:1$
mix\cite{beacom}.

Incidentally, neutrino decay also affects the signals for relic supernova 
$\bar{\nu}_e's$ and the sensitivity extends to $10^{10}$  sec/eV.
The main results can be summarized as follows\cite{fogli,ando}.  If we assume complete decay 
as before (for simplicity), then for normal hierarchy, the signal is enhanced
by about a factor of 2; and for inverted hierarchy, the signal goes away.

\section{Magnetic Moments and Other Neutrino Properties}

If the path of neutrinos takes them thru regions with significant magnetic 
fields and the neutrino magnetic moments are large enough, the flavor mix can 
be affected\cite{enquist}.  The main effect of the passage thru magnetic field is the 
conversion of a given helicity into an equal mixture of both helicity states.
This is also true in passage thru random magnetic fields\cite{domokos}.

If the neutrino are Dirac particles, and all magnetic moments are comparable, 
then the effect of the spin-flip is to simply reduce the overall flux of all 
flavors by half, the other half becoming the sterile Dirac partners.

If the neutrinos are Majorana particles, 
the flavor composition remains 1 : 1 : 1 when it 
starts from 1 : 1 : 1, and the absolute flux remains unchanged.

What happens when large magnetic fields are present in or near the neutrino 
production region?  In case of Dirac neutrinos, there is no difference and 
the outcoming flavor ratio remains 1 : 1 : 1, with the absolute fluxes
reduced by half.  In case of Majorana neutrinos,
since the initial flavor mix is no longer universal but is 
$\nu_e: \nu_\mu: \nu_\tau \approx 1: 2: 0,$ this is modified
but it turns out that the final(post-oscillation) flavor mix is still 1 : 1
: 1  !

As for mixing with sterile neutrinos, if the mixings are small, there are 
small deviations from the universality\cite{athar}.  A specific case of large mixing and 
very small $\delta m^2$ is discussed in the next section.

Other neutrino properties can also affect the neutrino flavor mix and modify
it from the canonical 1 : 1 : 1. If neutrinos have flavor(and equivalence
principle) violating couplings to gravity(FVG), or Lorentz invariance
violating(CPT violating or conserving) couplings; then there can be resonance
effects which make for one way transitions(analogues of MSW transitions)
e.g. $\nu_\mu \rightarrow \nu_\tau$ but not vice
versa\cite{minakata,barger3}. In case of FVG for example,
this can give rise to an anisotropic deviation of the $\nu_\mu/\nu_\tau$
ratio from 1, becoming less than 1 for events coming from the direction
towards the Great
Attractor, while remaining 1 in other directions\cite{minakata}. 
 
Another  possibility
that can give rise to deviations of the flavor mix from the canonical
1 : 1 : 1 is the idea of neutrinos of varying mass(MaVaNs). In this
proposal\cite{fardon}, by having the dark energy and neutrinos(a sterile one to be
specific) couple, and track each other; it is possible to relate the
small scale $2\times 10^{-3}$ eV required for the dark energy to the small
neutrino mass, and furthermore the neutrino mass depends inversely
on neutrino density, and hence on the epoch. As a result, if this
sterile neutrino mixes with a flavor neutrino, the mass difference
varies along the path, with potential resonance enhancement of the
transition probability into the sterile neutrino, and thus change the
flavor mix\cite{hung}. For example, if only one 
resonance is crossed enroute, it can 
lead to a conversion of the lightest (mostly) 
flavor state into the (mostly) sterile
state, thus changing the flavor mix to
 $1-U_{e1} ^2 \ : 1-U_{\mu 1}^2 \ : 
1-U_{\tau 1}^2
\approx 1/3 \ : 1 \ : 1,$ in case of normal 
hierarchy and similarly $\approx 2 \ : 1 \ : 1$
in case of inverted hierarchy.

\section{Pseudo-Dirac Neutrinos with very small mass differences
\cite{beacom3}}

If each of the three neutrino mass eigenstates is actually a doublet 
with very small mass difference (smaller than $10^{-6} eV)$, 
then there are no current experiments  that could have detected this. 
Such a possibility was raised long ago\cite{bilenky1}. 
It turns out that the only way to detect such small mass 
differences $(10^{-12} eV^2 > \delta m^2 > 10^{-18} eV^2)$ 
is by measuring flavor mixes of the high energy neutrinos 
from cosmic sources. Fig. 1 shows that relic supernova neutrino
signals and AGN neutrinos are sensitive to mass difference 
squared down to $10^{-20} eV^2$.

Let $(\nu_1^+, \nu_2^+, \nu_3^+; \nu_1^- \nu_2^-, \nu_3^-)$ 
denote the six mass eigenstates where $\nu^+$ and $\nu^-$ are a 
nearly degenerate pair.  A 6x6 mixing matrix rotates the mass 
basis into the flavor basis 
$(\nu_e, \nu_\mu, \nu_\tau; \nu_e^' \nu_\mu^', \nu_\tau^')$.  
In general, for six Majorana neutrinos, there would be fifteen 
rotation angles and fifteen phases.  However, for pseudo-Dirac 
neutrinos, Kobayashi and Lim\cite{kobayashi} have given an elegant proof 
that the 6x6 matrix $V_{KL}$ takes the very simple form 
(to lowest order in $\delta m^2 / m^2$:
\begin{eqnarray}
V_{KL} = \left (
\begin{array}{cc}
U & 0 \\
0 & U_R 
\end{array} \right) \cdot
\left (
\begin{array}{cc}
V_1 & iV_1 \\
V_2  & -iV_2 
\end{array}\right),
\end{eqnarray}
where the $3\times 3$ matrix U is just the usual mixing 
matrix determined by the atmospheric and solar observations, the 
$3\times 3$ matrix $U_R$ is an unknown unitary matrix and $V_1$ and 
$V_2$ are the diagonal matrices 
$V_1 =$ diag $(1,1,1)/\sqrt{2}$, and $V_2$=diag$(e^{-i \phi_1}, 
e^{-i \phi_2}, e^{-i \phi_3})/\sqrt{2}$, with the $\phi_i$ 
being arbitrary phases.

\begin{figure}[t]
\begin{center}
\centerline{\epsfxsize=0.5\textwidth\epsfbox{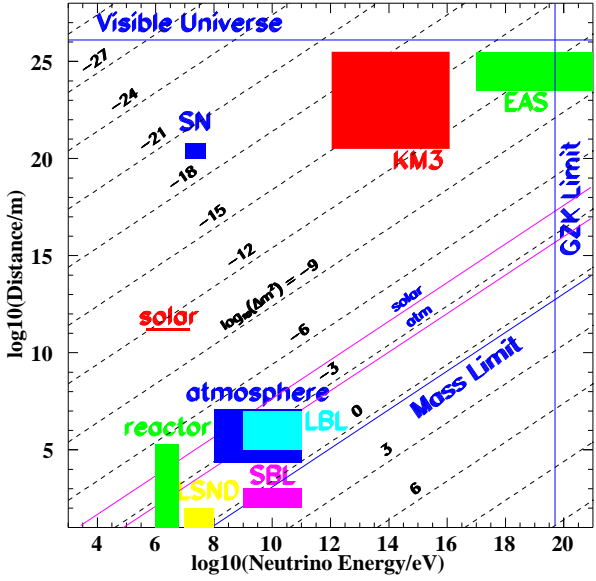}}
\caption{\label{learnedplot} The ranges of distance and energy covered
in various neutrino experiments.  The diagonal lines indicate the
mass-squared differences (in eV$^2$) that can be probed with vacuum
oscillations; at a given $L/E$, larger $\delta m^2$ values can be
probed by averaged-out oscillations.  We focus on a neutrino telescope
of 1-km scale (denoted ``KM3''), or larger, if necessary. From Ref.[28].}
\end{center}
\end{figure}

As a result, the three active neutrino states are described in terms of the six mass eigenstates as:
\begin{equation}
\nu_{\alpha L} = U_{\alpha j} \ \frac{1}{\sqrt{2}} \left(\nu^+_{j} + i \nu^-_{j}  \right).
\end{equation}

The nontrivial matrices $U_R$ and $V_2$ are not accessible to active flavor measurements.  The flavor conversion probability can thus be expressed as
\begin{eqnarray*}
P_{\alpha \beta} = \frac{1}{4} \left| \sum^{3}_{j=1} U_{\alpha j} 
\left\{ e^{i(m_j^+)^{2} L/2E} + e^{i (m_j^-)^{2} L/2E} \right \}
U^*_{\beta j} \right|^2 
\end{eqnarray*}

The flavor-conserving probability is also given by this formula, with $\beta =\alpha$.  Hence, in the description of the three active neutrinos, the only new parameters are the three pseudo-Dirac mass differences, 
$\delta m^2_j =  \left (m_j^+ \right )^2 - \left (m_j^- \right )^2$.
In the limit that they are negligible, 
the oscillation formulas reduce to the standard ones and there is no way to discern the pseudo-Dirac nature of the neutrinos.

{\bf $L/E$-Dependent Flavor Ratios.---}
Given the enormous pathlength between astrophysical neutrino sources
and the Earth, the phases due to the relatively large solar and
atmospheric mass-squared differences will average out (or
equivalently, decohere). The probability for a neutrino telescope to measure
the flavor $\nu_\beta$ is then:
\begin{eqnarray*}
\label{prob4}
P_\beta = \sum_\alpha w_\alpha \sum_{j=1}^3
|U_{\alpha j}|^2\, |U_{\beta j}|^2 \,
\left[1-\sin^2\left(\frac{\delta m^2_j\,L}{4E}\right)\right]
\end{eqnarray*}
where $w_\alpha$ represents the fraction of the flavor $\alpha$
present initially.
In the limit that $\delta m^2_j \rightarrow 0$, the expression above
reproduces the standard form.  The new oscillation terms are
negligible until $E/L$ becomes as small as the tiny pseudo-Dirac
mass-squared splittings $\delta m^2_j$.

\begin{table}[t]
\caption{\label{ratios} Flavor ratios $\nu_e:\nu_\mu$ for various
scenarios.  The numbers $j$ under the arrows denote the pseudo-Dirac
splittings, $\delta m^2_j$, which become accessible as $L/E$
increases.  Oscillation averaging is assumed after each transition
$j$.  We have used $\theta_{\rm atm} = 45^\circ$, $\theta_{\rm solar}
= 30^\circ$, and $U_{e3} = 0$.}
\begin{center}

\begin{tabular}{ccccccc}\hline\hline
$1:1$ & $_{\stackrel{\longrightarrow}
{3}}$ & 4/3:1 & $_{\stackrel{\longrightarrow}
{2,3}}$ & $14/9:1$ &
$_{\stackrel{\longrightarrow}{1,2,3}}$ & $1:1$  \\
$1:1$ & $_{\stackrel{\longrightarrow}
{1}}$ & 2/3:1 & $_{\stackrel{\longrightarrow}
{1,2}}$ & $2/3:1$ &

$_{\stackrel{\longrightarrow}{1,2,3}}$ & $1:1$  \\
$1:1$ & $_{\stackrel{\longrightarrow}
{2}}$ & 14/13:1 & $_{\stackrel{\longrightarrow}
{2,3}}$ & $14/9:1$ &
$_{\stackrel{\longrightarrow}{1,2,3}}$ & $1:1$  \\
$1:1$ & $_{\stackrel{\longrightarrow}
{1}}$ & 2/3:1 & $_{\stackrel{\longrightarrow}
{1,3}}$ & $10/11:1$ &
$_{\stackrel{\longrightarrow}{1,2,3}}$ & $1:1$  \\
$1:1$ & $_{\stackrel{\longrightarrow}
{3}}$ & 4/3:1 & $_{\stackrel{\longrightarrow}
{1,3}}$ & $10/11:1$ &
$_{\stackrel{\longrightarrow}{1,2,3}}$ & $1:1$  \\
$1:1$ & $_{\stackrel{\longrightarrow}
{2}}$ & 14/13:1 & $_{\stackrel{\longrightarrow}
{1,2}}$ & $2/3:1$ &
$_{\stackrel{\longrightarrow}{1,2,3}}$ & $1:1$  \\
\hline\hline
\end{tabular}
\end{center}
\end{table}

The flavors deviate from the democratic value of $\third$ by
\begin{eqnarray*}
\label{deviate3}
\delta P_e &=& -\third\,\left[ \frac{3}{4}\chi_1 + \quarter\,
\chi_2 \right],\nonumber\\
\delta P_\mu = \delta P_\tau  &=& -\third\,\left[ 
\frac{1}{8}\chi_1 + \frac{3}{8}\,\chi_2 
	+\half\,\chi_3\right] \,
\end{eqnarray*}
where $\chi_i = \sin^2(\delta m_i^2 L/4E)$.

Table~\ref{ratios} shows how the $\nu_e : \nu_\mu$ ratio is
altered if we cross the threshold for one, two, or all three of the
pseudo-Dirac oscillations.  The flavor ratios deviate from $1:1$ when
one or two of the pseudo-Dirac oscillation modes is accessible.  In
the ultimate limit where $L/E$ is so large that all three oscillating
factors have averaged to $\half$, the flavor ratios return to $1:1$,
with only a net suppression of the measurable flux, by a factor of
$1/2$.

\section{\bf Cosmology with Neutrinos} 
If the oscillation phases can indeed be measured for the very small 
mass differences by the deviations of the flavor mix from 1 : 1 : 1 as 
discussed above, the
following possibility is raised. It is a fascinating fact 
that non-averaged oscillation phases,
$\delta\phi_j=\delta m_j^2 t/4p$, and hence the factors $\chi_j$, are
rich in cosmological information\cite{weiler,weiler1}.  
Integrating the phase
backwards in propagation time, with the momentum blue-shifted, 
one obtains
\begin{eqnarray}
\delta\phi_j&=&\int_0^{z_e} dz\frac{dt}{dz}\frac{\delta m_j^2}{4p_0(1+z)}\\
	  &=&\left(\frac{\delta m_j^2 H^{-1}_0}{4p_0}\right) \ I
\end{eqnarray}
where I is given by
\begin{equation}
I=\int_1^{1+{z_e}}\frac{d\omega}{\omega^2}
		\frac{1}{\sqrt{\omega^3\,\Omega_m+(1-\Omega_m)}}\,,\nonumber
\end{equation}
$z_e$ is the red-shift of the emitting source, and $H_0^{-1}$ is
the Hubble time, known to 10\%~\cite{freedman}.  
This result holds
for a flat universe, where 
$\Omega_m+\Omega_\Lambda=1$, with $\Omega_m$ and 
$\Omega_\Lambda$ the matter and vacuum energy
densities in units of the critical density.
The integral $I$
is the fraction of the Hubble time 
available for neutrino transit.
For the presently preferred values $\Omega_m=0.3$ and
$\Omega_\Lambda=0.7$, the asymptotic ($z_e\rightarrow\infty$) value of
the integral is 0.53.  This limit is approached rapidly: at
$z_e=1\,(2)$ the integral is 
77\% (91\%) saturated.  
For cosmologically distant ($z_e > 1$) sources such as gamma-ray bursts, 
non-averaged oscillation data would, in principle, allow one to
deduce $\delta m^2$ to about 20\%, without even knowing the source
red-shifts. Known values of $\Omega_m$ and $\Omega_\Lambda$ might 
allow one to infer the source redshifts $z_e$, or vice-versa.

This would be the first measurement of a cosmological
parameter with particles other than photons.  An advantage of
measuring cosmological parameters with neutrinos is the fact that
flavor mixing is a microscopic phenomena and hence presumably free of
ambiguities such as source evolution or standard candle
assumptions\cite{weiler,stodolsky}.  Another method of
measuring cosmological parameters with neutrinos is given in
Ref.\cite{choubey}.

\section{Experimental Flavor Identification} 

It is obvious from the above discussion that flavor identification is
crucial for the purpose at hand. In a water cerenkov detector flavors can be 
identified as follows.

The $\nu_\mu$ flux can be measured by the $\mu's$ produced by the charged 
current interactions and the resulting $\mu$ tracks in the detector which
are long  at these energies.  $\nu_{e}'{s}$ produce showers by both
CC and NC interactions.  The total rate for showers includes those
produced by NC interactions of $\nu_\mu's$ and $\nu_\tau's$ as well and those
have to be (and can be) subtracted off to get the real flux of $\nu_e's$.
However, this distinction between hadronic showers of neutral current
events and the eletron-containing charged current events is rather 
difficult to make. Double-bang and lollipop events are signatures unique to 
tau neutrinos, made
possible by the fact that tau leptons decay before they lose a significant
fraction of their energy.  Double-bang events consists of a hadronic shower
initiated by a charged-current interaction of the $\nu_\tau$ followed by a
second energetic shower (hadronic or electromagnetic) from the decay  of the
resulting tau lepton\cite{learned1}.  Lollipop events consist of the second
of
the double-bang showers along with the reconstructed tau lepton track (the
first bang may be detected or not).  In principle, with a sufficient number
of
events, a fairly good estimate of the flavor ratio $\nu_e: \nu_\mu: \nu_\tau$ 
can be reconstructed, as has been discussed recently.
Deviations of the flavor ratios 
from $1:1:1$ due to possible decays are so extreme that they should be
readily identifiable\cite{beacom4}. Upcoming high energy neutrino telescopes,
such as Icecube\cite{karle}, will not have perfect ability to separately 
measure the neutrino flux in each flavor.  However, the quantities 
we need are closely related to 
observables, in particular in the limit of $\nu_\mu - \nu_\tau$ symmetry 
$(\theta_{atm} = 45^0$ and $U_{e3} = 0)$, in which all mass 
eigenstates contain equal fractions of $\nu_\mu$ and 
$\nu_\tau$.  In that limit, the fluxes for $\nu_\mu$ and 
$\nu_\tau$ are always in the ratio 1 : 1, with or without decay. 
This is useful since the $\nu_\tau$ flux is the hardest to measure. 

Even in the extreme case when one assumes that tau events are not 
identifiable, something about the flavor mix can be deduced. Let the only 
experimental information available be the number of muon tracks and the 
number of showers.The relative number of shower events to track events can 
be related to the most interesting quantity for testing decay scenarios, 
i.e., the $\nu_e$ to $\nu_\mu$ ratio.  The precision of the upcoming 
experiments should be good enough to test the extreme flavor ratios produced
by decays.  If electromagnetic and hadronic  showers can be separated, then 
the precision will be even better\cite{beacom4}.

Comparing, for example, the standard flavor ratios of 1 : 1 : 1 to the
possible 5 : 1 : 1 generated by decay, the more numerous electron neutrino
flux will result in a substantial increase in the relative number of 
shower events.

The details of this observation depends on the range of muons generated in
or around the detector and the ratio of charged to neutral current cross
sections.  The measurement will be limited by the energy resolution of the
detector and the ability to reduce the atmospheric neutrino background.  The
atmospheric background drops rapidly with energy and should be negligibly small at and above the PeV scale.

\section{Discussion and Conclusions}  

The flux ratios we discuss are energy-independent because we have assumed that the ratios at production are energy-independent, that all oscillations are averaged out, and that all possible decays are complete.  In the standard scenario with only oscillations, the final flux ratios are $\phi_{\nu_{e}} :  \phi_{\nu_{\mu}} : 
 \phi_{\nu_{\tau}} = 1 : 1 : 1$.  In the cases with decay, we have found rather
different possible flux ratios, for example 5 : 1 : 1 in the normal hierarchy and 
0 : 1 : 1 in the inverted hierarchy.  These deviations from 1 : 1 : 1 are so extreme that they should be readily measurable.

If we are very fortunate\cite{barenboin}, we may be able to observe a reasonable number of events from several sources (of known distance) and/or over a sufficient range in energy.  Then the resulting dependence of the flux ratio 
$(\nu_e/\nu_\mu)$ on L/E as it evolves from say 5 (or 0) to 1, can be clear
evidence of decay and further can pin down the actual lifetime instead of
just placing a bound.

To summarize, we suggest that if future measurements of the flavor mix at
earth of high energy astrophysical neutrinos find it to be
\begin{equation}
\phi_{\nu_{e}} / \phi_{\nu_{\mu}} / \phi_{\nu_{\tau}} = \alpha / 1 / 1 ;
\end{equation}
then
\begin{description}
\item[(i)] $\alpha \approx 1$ (the most boring case) confirms our knowledge of the
MNSP\cite{MNS} matrix and our prejudice about the production mechanism;
\item[(ii)] $\alpha \approx 1/2$ indicates that the source emits pure
$\nu_\mu's$ and the mixing is conventional;
\item[(iii)]$\alpha \approx 3$ from a unique direction, e.g. the Cygnus region, would be
evidence in favour of a pure $\bar{\nu}_e$ production as has been suggested
recently\cite{goldberg};
\item[(iv)] $\alpha > 1$ indicates that neutrinos are decaying with normal
hierarchy; and 
\item[(v)]$\alpha \ll 1$ would mean that neutrino decays are occuring with
inverted hierarchy;
\item[(vi)] Values of $\alpha$ which cover a broader range (3 to 15) and 
deviation of the $\mu/\tau$ ratio from 1(between 0.2 to 5) can yield valuable 
information about $U_{e3}$ and $\cos \delta$. Deviations of $\alpha$ which are less extreme(between 0.7 and 1.5) can also probe very small pseudo-Dirac 
$\delta m^2$ (smaller than $10^{-12} eV^2$).
\end{description}  

Incidentally, in the last three cases, the results have absolutely no
dependence on the initial flavor mix, and so are completely free of any
dependence on the production model. So either one learns about the production
mechanism and the initial flavor mix, as in the first three cases, or one
learns only about the neutrino properties, as in the last three cases.
In any case, it should be evident that the construction of very large neutrino detectors is a ``no lose'' proposition.

\section{Acknowledgements}
This talk is based on published and ongoing work in collaboration with
John Beacom, Nicole Bell, Dan Hooper, John Learned  and Tom Weiler. I thank them for a most
enjoyable collaboration; I also thank them for a
careful and critical reading of the
manuscript. I would like to acknowledge the splendid
hospitality of the Fujihara foundation and Yoji Totsuka and Kenzo Nakamura
in Tsukuba.
This work was supported in part by U.S.D.O.E. under grant DE-FG03-94ER40833.


\begin{thebibliography}{99}

\bibitem{pakvasa}S. Pakvasa, 9th International Symposium 
on Neutrino Telescopes, Venice, Italy, 6-9 Mar 2001, Venice 2001, Neutrino 
Telescopes, ed. M. Baldo-Ceolin, Vol. 2, p. 603; hep-ph/0105127.

\bibitem{MNS}Z. Maki, M. Nakagawa and S. Sakata, {\it Prog. Theoret. Phys.} {\bf 28}, 870 (1962); see also V. N. Gribov and B. M. Pontecorvo, {\it Phys. Lett.}
{\bf B28}, 493 (1969); the generalization to three flavors first
appeared in B. W. Lee, S. Pakvasa, R. Shrock and H. Sugawara, {\it Phys. Rev. Lett.} {\bf 38}, 937 (1977).

\bibitem{learned}J. G. Learned and K. Mannheim, 
{\it Ann. Rev. Nucl. Part. Sci.} {\bf 50}, 603 (2000), 
and references therein.

\bibitem{learned1}J. G. Learned and S. Pakvasa, 
{\it Astropart. Phys.} {\bf 3},
267 (1995); hep-ph/9405296.

\bibitem{rachen}J. P. Rachen and P. Meszaros, {\it Phys. Rev.} {\bf D58}, 
123005 (1998); astro-ph/9802280.

\bibitem{pakvasa1}S. Pakvasa and J. Valle,  {\it Proc. Indian. Natl. Sci. Acad.}
{\bf 70A}, 189 (2003); hep-ph/0301061.

\bibitem{athar}H. Athar, M. Jezabek and O. Yasuda, {\it Phys. Rev.} 
{\bf D62}, 103007 (2000); hep-ph/0005104; L. Bento, P. Keranen and J. Maalampi,
{\it Phys. Lett.} {\bf B476}, 205 (2000); hep-ph/9912240.


\bibitem{beacom}J. F. Beacom, N. Bell, D. Hooper, S. Pakvasa and T.J. Weiler,
{\it Phys. Rev. Lett.} {\bf 91}, 181301 (2003); hep-ph/0211305.

\bibitem{pakvasa2}S. Pakvasa, {\it Physics Potential and Development of Muon Colliders}, Mu 99, San Francisco, AIP Conf. Proc. {\bf 542} (2000) 99, ed D. Cline, p. 99; hep-ph/0004077.

\bibitem{john}J. F.  Beacom and P. Vogel, {\it Phys. Rev. Lett.} {\bf 83}, 5222
(1999); hep-ph/9970383.

\bibitem{these}These are deduced\cite{pakvasa2} from the current  bounds on the magnetic moments: C. Caso et al., Review of Particle Properties, Particle Data Group, 
{\it Eur. Phys. J.} {\bf C3}, 1 (1998).

\bibitem{frere}J-M Frere, R. B. Nezorov and M. Vysotsky, {\it Phys. Lett.}
{\bf B394}, 127 (1997); hep-ph/9608266.

\bibitem{bilenky}M. Bilenky and A. Santamaria, hep-ph/9908272; 
{\it Phys. Lett.} {\bf B301}, 287 (1993).

\bibitem{gelmini}G. Gelmini and M. Roncadelli, {\it Phys. Lett.} {\bf B99},
411 (1981).

\bibitem{chicasige}Y. Chikasige, R. Mohapatra and R. Peccei, {\it Phys. Rev. Lett.} {\bf 45}, 1926 (1980).

\bibitem{valle}J. Valle, {\it Phys. Lett.}, {\bf B131}, 87 (1983); G. Gelmini
and J. Valle, {\it ibid}. {\bf B142}, 181 (1983); A. Joshipura and S. Rindani, {\it Phys.
Rev.} {\bf D46}, 3008 (1992); A. Acker, A. Joshipura and S. Pakvasa, {\it Phys. Lett.} {\bf B285}, 371 (1992); A. Acker, S. Pakvasa and J. Pantaleone, {\it Phys. Rev.} {\bf D45}, 1  (1992).

\bibitem{choi}A. Choi and A. Santamaria, {\it Phys.Lett.} {\bf B267}, 504 (1991).

\bibitem{barger}V. Barger, W-Y. Keung and S. Pakvasa, {\it Phys. Rev.} {\bf D25}, 907 (1982).

\bibitem{beacom1}J.F. Beacom, N. Bell and S. Doddelson, hep-ph/0404585. 

\bibitem{chacko} Z. Chacko, L.J. Hall, T. Okui and S.J. Oliver, hep-ph/0312267.

\bibitem{hirata} K. Hirata et al., {\it Phys. Rev. Lett.} {\bf 58}, 1497
(1988); R.M. Bionta et al., ibid. 58, 1494 (1988).

\bibitem{eguchi}K. Eguchi et al; {\it Phys. Rev. Lett.} {\bf 92}, 071301 
(2004); hep-ex/0310047.

\bibitem{bell}J. F. Beacom and N. Bell; {\it Phys. Rev.} {\bf D65}, 113009
(2002); hep-ph/0204111; and references cited therein.

\bibitem{barger1}V. D. Barger, J. G. Learned, S. Pakvasa and T. J. Weiler, {\it Phys. Rev. Lett.} {\bf 82}. 2640 (1999); hep-ph/9810121; Y. Ashie et al., hep-ex/0404034.

\bibitem{pakvasa3}S. Pakvasa, {\it Lett. Nuov. Cimm.} {\bf 31}, 497 (1981); Y. Farzan and A. Smirnov, {\it Phys. Rev.} {\bf D65}, 113001 (2002); hep-ph/0201105.

\bibitem{weiler}T. J. Weiler, W. A. Simmons, S. Pakvasa and J. G. Learned; hep-ph/9411432.

\bibitem{keranen}P. Keranen, J. Maalampi and J. T. Peltonieni, {\it Phys. Lett.} {\bf B461}, 230 (1999); hep-ph/9901403.


\bibitem{lindner}M. Lindner, T. Ohlsson and W. Winter, {\it Nucl. Phys.} 
{\bf B607}, 326 (2001); ibid, {\bf B622}, 429 (2002); astro-ph/0105309.

\bibitem{beacom2}J. F. Beacom, N. Bell, D. Hooper, S. Pakvasa and T. J.
Weiler, {\it Phys. Rev} {\bf D69}, 017303 (2004); hep-ph/0309267.

\bibitem{fogli}G. L. Fogli, E. Lisi, A. Minzi and D. Montanino,
hep-ph/0401227.

\bibitem{ando}S. Ando, {\it Phys. Lett.} {\bf B570}, 11 (2003); hep-ph/0307169.

\bibitem{enquist} K. Enqvist, P. Keranen and J. Maalampi, {\it Phys. Lett.}{\bf B438},295(1998); hep-ph/9806392.

\bibitem{domokos}G. Domokos and S. Kovesi-Domokos, {\it Phys. Lett} {\bf B410}, 57 (1997); hep-ph/9703265.

\bibitem{minakata}H. Minakata and A. Yu. Smirnov, {\it Phys. Rev.} {\bf D54}, 
3698 (1996); hep-ph/9601311.

\bibitem{barger3}V. D. Barger, S. Pakvasa, T. J. Weiler and K. Whisnant, {\it Phys. Rev. Lett.} {\bf 85}, 5055 (2000); hep-ph/0005197.

\bibitem{fardon}R. Fardon, A. E. Nelson and N. Weiner, astro-ph/0309800.

\bibitem{hung}P. Q. Hung and H. Pas, astro-ph/0311131.
 
\bibitem{beacom3}J. F. Beacom, N. Bell, D. Hooper, J. G. Learned, S. Pakvasa 
and T. J. Weiler; {\it Phys. Rev. Lett.}, {\bf 92} (2004); hep-ph/0307151; 
see also P. Keranen, J. Maalampi, M. Myyrylainen and J. Riittinen,
{\it Phys. Lett.} {\bf B574}, 162 (2003); hep-ph/0307041 for similar 
considerations.

\bibitem{bilenky1}L. Wolfenstein, {\it Nucl. Phys.} {\bf B186}, 147 (1981);
S. M. Bilenky and B. M. Pontecorvo, {\it Sov. J. Nucl. Phys.} {\bf 38}, 248 (1983); S.T. Petcov, 
{\it Phys. Lett.} {\bf B110}, 245 (1982).

\bibitem{kobayashi} M. Kobayashi and C. S. Lim, {\it Phys. Rev.} {\bf D64}, 013003 (2001); hep-ph/0012266.

\bibitem{weiler1}D. J. Wagner and T. J. Weiler, {\it Mod. Phys. Lett.} {\bf A12}, 2497 (1997).

\bibitem{freedman}W. L. Freedman et al. {\it Astrophys. J.} {\bf 553}, 47 (2001).

\bibitem{stodolsky}L. Stodolsky, {\it Phys. Lett.} {\bf B473}, 61 (2000); astro-ph/9911167.

\bibitem{choubey}S. Choubey and S. F. King, {\it Phys. Rev.} {\bf D67}, 073005 (2003), hep-ph/0207260.


\bibitem{beacom4}J. F. Beacom, N. Bell, D. Hooper, S. Pakvasa and T. J.
Weiler; {\it Phys. Rev.} {\bf D68}, 093005 (2003); hep-ph/0307025; F. Halzen and D. Hooper, {\it Rept. Prog. Phys.} 
{\bf 65}, 1025 (2002); astro-ph/0204527.

\bibitem{karle}A. Karle, {\it Nucl. Phys. Proc. Supp.}, {\bf 118} (2003);
astro-ph/0209556; A. Goldschmidt, {\it Nucl. Phys. Proc. Suppl.} {\bf 110},
516 (2002).


\bibitem{barenboin}G. Barenboim and C. Quigg, {\it Phys. Rev.} {\bf D67}, 073024 (2003); hep-ph/0301220.

\bibitem{goldberg}L. A. Anchordoqui, H. Goldberg, F. Halzen and T. J. Weiler,
astro-ph/0311002.
\end{thebibliography}
\end{document}